\documentclass[aps,preprint]{revtex4}
\usepackage{eurosym}
\usepackage{amsfonts}
\usepackage{amsmath}
\usepackage{amssymb}

\begin{document}

\title{K-essence sources of Kerr--Schild spacetimes}
\author{Bence Juh\'{a}sz$^{1,2}$}
\author{L\'{a}szl\'{o} \'{A}rp\'{a}d Gergely$^{1,2}$}
\affiliation{$^{1}$Department of Theoretical Physics, University of Szeged, Tisza Lajos
krt. 84-86, H-6720 Szeged, Hungary}
\affiliation{\thinspace $^{2}$Department of Theoretical Physics, HUN-REN Wigner Research
Centre for Physics, Konkoly-Thege Mikl\'{o}s \'{u}t 29-33, H-1121 Budapest,
Hungary}
\date{\today }

\begin{abstract}
We extend a result by one of the authors, established for nonvacuum Einstein
gravity, to minimally coupled k-essence scalar-tensor theories. First we
prove that in order to source a Kerr--Schild type spacetime, the k-essence
Lagrangian should be at most quadratic in the kinetic term. This is reduced
to linear dependence when the Kerr--Schild null congruence is autoparallel.
Finally, we show that requiring the solutions of the Einstein equations
linearized in Kerr--Schild type perturbations to also solve the full
nonlinear system of Einstein equations, selects once again k-essence scalar
fields with Lagrangians linear in the kinetic term. The only other k-essence
sharing the property of sourcing perturbative Kerr--Schild spacetimes which
are also exact, is the scalar field constant along the integral curves of
the Kerr--Schild congruence, with otherwise unrestricted Lagrangian.
\end{abstract}

\maketitle

\section{Introduction}

Most physically interesting metrics in Einstein gravity are of Kerr--Schild
type. They include Schwarzschild and Kerr black holes for vacuum, also the
Kerr--Newman family and pp-waves for Einstein--Maxwell systems or the Vaidya
radiating solution sourced by a null dust \cite{ExactSol}. Such spacetimes
are generated by a null congruence $l^{a}$ through the map 
\begin{equation}
\tilde{g}_{ab}=g_{ab}+\lambda l_{a}l_{b}  \label{KS}
\end{equation}%
from the flat metric $g_{ab}=\eta _{ab}$, with $\lambda $ an arbitrary
parameter. The extension to a generic vacuum seed metric $g_{ab}$ led to
either a shearfree congruence $l^{a}$ (containing all solutions with flat
seed metric) or a unicity theorem for the shearing class (only containing
one of the K\'{o}ta--Perj\'{e}s metrics and its nontwisting limit, the
Kasner metric) \cite{KS1,KS2,KSB0}. An important result was provided by
Xanthopoulos \cite{Xanthopoulos}, stating that all vacuum Kerr--Schild
metrics arising as perturbations (with small $\lambda $) of vacuum seed
spacetimes are also exact (hence solutions of the Einstein equations for
arbitrary $\lambda $). This result was generalized for the nonvacuum case by
one of us \cite{GAL}, proving that for any pair $\left( g_{ab},T_{ab}\right) 
$ of seed metric and energy-momentum tensor, the pair $\left( \tilde{g}%
_{ab},T_{ab}+\lambda T_{ab}^{\left( 1\right) }\right) $ arising as solution
of the linearized solution (hence for small $\lambda $) generates an exact
solution (with arbitrary $\lambda $) of the form $\left( \tilde{g}%
_{ab},T_{ab}+\lambda T_{ab}^{\left( 1\right) }+\lambda
^{2}l_{(a}T_{b)c}^{\left( 1\right) }l^{c}\right) $, in the case when the
null congruence is autoparallel (if not, a similar, but more technical
result holds).

While general relativity is precisely verified by Solar System tests, also
all cosmological, astrophysical and gravitational wave observations are
consistent with it, modified gravitational theories are still of interest,
provided they obey the observational constraints. They are pursued driven by
the need to explain dark matter and dark energy, to properly model inflation
and make room for quantum gravity motivated extensions in the low energy
regime. A plethora of modifications relaxing one or more assumptions of the
Lovelock theorem (gravity expressed solely by the metric tensor, obeying
second order dynamics in 4 spacetime dimensions with a divergence-free
energy-momentum tensor) are still viable \cite{TestingGRtopical}. The
relation of some of these modified theories with a subclass of Kerr--Schild
metrics has been investigated for vector-tensor \cite{GursesVT,GursesH}
theories.

A most natural modification arises by including a scalar field into the
gravitational sector. By imposing second order dynamics for both the metric
and the scalar, such that Ostrogradsky instabilities are avoided, the
Horndeski class emerges \cite{Horndeski,Deffayet}. Additionally, requiring
gravitational waves to propagate with the speed of light in vacuum (in order
to comply with observations of high frequency gravitational waves by LIGO
and Virgo \cite{Multimessenger}), leads to \cite{GWc1,GWc2,GWc3,GWc4} a
restricted subclass of kinetic gravity braiding theories \cite{KGB}. The
dependence of the Lagrangian on the scalar $\phi $ of such theories is only
through $\phi ,~\square \phi $ and the kinetic term 
\begin{equation}
X=-\dfrac{1}{2}g^{ab}\nabla _{a}\phi \nabla _{b}\phi \ 
\end{equation}%
(with $g^{ab}$ the inverse metric, $\nabla _{a}$ the Levi-Civita connection
and $\square =\nabla _{a}\nabla ^{a}$). Without the $\square \phi $
dependence, they simplify to the k-essence class of scalar-tensor theories
with Lagrangian%
\begin{equation}
\mathfrak{L}_{\phi }=\sqrt{-\mathfrak{g}}F(\phi ,X)  \label{l2}
\end{equation}%
(here $\mathfrak{g}$ is the determinant of the metric and $F(\phi ,X)$ an
arbitrary function of $\phi $ and $X$). We will further assume minimal
coupling to the metric, such that $\mathfrak{L}_{\phi }$ is supplemented
with the Einstein--Hilbert action (in this case the Einstein and Jordan
frames coincide).

Such scalar dynamics was originally introduced for k-inflation models, with
Lagrangians combining first and second order powers of $X$, and explored in
the context of slow-roll and power law inflation scenarios \cite{kinfl}. A
k-essence model with the Lagrangian consisting of a purely $X$-dependent
function divided by the scalar squared was proposed to generate late time
dark energy through the transformation of the scalar field into a negative
pressure state \cite{lateDE}. A related k-essence Lagrangian consisting of a
product of functions depending solely of $\phi $ and $X$, respectively is
suitable to accommodate for slow-roll, power-law and pole-like inflation
mechanisms, and it also appears in the effective action of string theory 
\cite{kess}. Such models are able to generate the cosmic evolution and they
have the appealing feature that late time acceleration is not permanent.

Moreover, Lim, Sawicki, and Vikman proposed a unification of Dark Matter and
Dark Energy in a single degree of freedom \cite{DDE}. In their model, the
k-essence field is supplemented by a second scalar acting as a\ Lagrange
multiplier (hence, without a \ proper kinetic term), enforcing a relation
between $X$ and $\phi $. As consequence of this constraint, the system
retains a single dynamical degree of freedom, allowing for no wave-like
modes, hence leading to a generalised k-essence with vanishing speed of
sound, energy flowing along timelike geodesics (similarly as for dust),
while possessing non-zero pressure. This model is able to reproduce the $%
\Lambda $CDM evolution, with structure formation possibly affected. Similar
techniques are explored for predicting WIMP dark matter mass spectrum, while
the nonvanishing pressure of matter mimicks the cosmological constant \cite%
{Speeding}. Generalised k-essence was explored to heal the cosmological
constant problem in Ref. \cite{Healing}, also to achieve a graceful exit
from inflation through proper choices of the scalar potential \cite{Luongo}.

Gravitational collapse of k-essence was investigated numerically in Ref. 
\cite{KEBH}. For strong fields yielding to black hole formation, in certain
cases the sound horizon may lie inside the light horizon, allowing for
superluminal k-essence signals escaping the black hole. The evolution of a
k-essence scalar field is governed by an effective metric (different from
the spacetime metric). Its conformally related emergent gravity metric
resembled a generalized Vaidya metric sourced by a superposition of perfect
and null fluids, when the scalar was assumed to be driven by a
Dirac--Born--Infeld type dynamics and assumed to depend only on one of the
advanced and retarded null coordinates \cite{Manna1}. Gravitational collapse 
\cite{Manna2}\ and the evaporation of the emerging horizon \cite{Manna3}
were also discussed.

In the present paper we aim to investigate another feature related to
k-essence, namely, under which conditions would the linearity property of
Kerr--Schild metrics proven in Ref. \cite{GAL} apply in this class of
minimally coupled k-essence scalar fields, also dubbed Class (A) in the
classification of Ref. \cite{classific}. In Section 2 we summarize the
results of Ref. \cite{GAL} on Kerr--Schild spacetimes with matter sources,
necessary for our analysis on minimally coupled k-essence fields.

In Section 3 we impose the condition on the k-essence to source Kerr--Schild
spacetimes. In Section 4 we analyse the requirements for lifting the
solution of the linearized system to exact solution by increasing the
Kerr--Schild parameter to arbitrary values. In Section 5 we repeat the
analysis for a simpler case, left out from the previous discussion.

In Section 6 we address the question of how are black hole properties
affected by Kerr--Schild maps. We also include an analysis of the scalar
fields inside or outside the event horizons of black holes, also of
cosmological scalar fields, in terms of equivalent fluids. We calculate the
adiabatic speed of sound, which does not vanish for the types of k-essence
scalar fields allowed by our requirements, a property already noted in Ref. 
\cite{KEBH}, and we formally exclude the Laplacian instability regimes.
Finally, we argue that k-essence Kerr--Schild seed spacetimes could be
important in dynamical situations.

In Section 7 we summarize our results.

\section{Kerr--Schild spacetimes and k-essence}

In this section we summarize the main results of Ref. \cite{GAL} necessary
for our forthcoming discussion and discuss how the k-essence fits into the
generic scheme.

\subsection{Nonvacuum Kerr--Schild maps}

The Ricci tensors of the Kerr--Schild and seed spacetimes are related as:

\begin{equation}
\Tilde{R}_{ab}=R_{ab}+\lambda R_{ab}^{(1)}+\lambda ^{2}R_{ab}^{(2)}+\lambda
^{3}R_{ab}^{(3)},  \label{RicciExp}
\end{equation}%
with the contributions%
\begin{equation}
R_{ac}^{(1)}=\nabla _{b}\bigg(\nabla _{(a}\left( l_{c)}l^{b}\right) -\dfrac{1%
}{2}\nabla ^{b}\left( l_{a}l_{c}\right) \bigg),
\end{equation}

\begin{equation}
R_{ac}^{(2)}=\nabla_b l^b l_{(a} D l_{c)}+ \dfrac{1}{2}D l_a D l_c +
l_{(a}DD l_{c)} + l_a l_c \nabla_b l_d \nabla^{[b}l^{d]}-D l^b \nabla_b
l_{(a} l_{c)} \ ,
\end{equation}

\begin{equation}
R_{ac}^{(3)}=-\dfrac{1}{2}l_{a}l_{c}Dl^{b}Dl_{b}\ 
\end{equation}%
(here $Dl^{a}=l^{b}\nabla _{b}l^{a}$ is the directional covariant derivative
along the null congruence). Expressing the Ricci tensor contributions in Eq.
(\ref{RicciExp}) through the Einstein equations written for both the seed
and Kerr--Schild metrics, the condition 
\begin{equation}
\lambda R_{ab}^{(1)}+\lambda ^{2}R_{ab}^{(2)}+\lambda ^{3}R_{ab}^{(3)}=%
\Tilde{T}_{ab}-T_{ab}-\frac{1}{2}g_{ab}\left( \Tilde{T}-T\right) -\dfrac{1}{2%
}\lambda l_{a}l_{b}\Tilde{T}\   \label{feltetel}
\end{equation}%
emerges (we have absorbed the constants into a redefinition of the
energy-momentum tensors). In Ref. \cite{GAL} it was proven that when seeking
the source of the Kerr--Schild spacetime in the form of the series%
\begin{equation}
\Tilde{T}_{ab}=T_{ab}+\lambda T_{ab}^{(1)}+\lambda ^{2}T_{ab}^{(2)}+\lambda
^{3}T_{ab}^{(3)}+\sum_{i=1}^{\infty }\lambda ^{3+i}T_{ab}^{(3+i)}\ ,
\label{GEM}
\end{equation}%
the terms of higher order than three vanish. Furthermore, when $l^{a}$ is
autoparallel, $T_{ab}^{(3)}=0$ also holds. Additionally requiring that the
solution for small $\lambda $ (the solution of the linear equation) solves
the full set of Einstein equations, leads to the condition 
\begin{equation}
T_{ab}^{(2)}=l_{(a}T_{b)c}^{\left( 1\right) }l^{c}~.  \label{cond}
\end{equation}%
This was announced as Theorem 2 in Ref. \cite{GAL}.

\subsection{K-essence}

The dynamics of the k-essence is given by the action%
\begin{equation}
S_{\phi }=\int \text{d}^{4}x\mathfrak{L}_{\phi }=\int \text{d}^{4}x\sqrt{-%
\mathfrak{g}}F(\phi ,X)\ ,  \label{action}
\end{equation}%
while its energy-momentum tensor emerges from its metric variation, 
\begin{equation}
T_{ab}=\dfrac{-2}{\sqrt{-\mathfrak{g}}}\dfrac{\delta S_{\phi }}{\delta g^{ab}%
}\ 
\end{equation}%
as 
\begin{equation}
T_{ab}=F_{X}(\phi ,X)\nabla _{a}\phi \nabla _{b}\phi +g_{ab}F(\phi ,X)~.
\label{Tab}
\end{equation}%
Here the subscript $X$ denotes the derivative with respect to $X$.

As we assume minimal coupling, the variation of the total action, the sum of
the Einstein--Hilbert action and the k-essence contribution (\ref{action}),
with respect to the metric gives the Einstein equations sourced by the
energy-momentum tensor (\ref{Tab}). Hence, in this case the results of Ref. 
\cite{GAL} can be applied directly.

While the k-essence field is unaffected by the Kerr--Schild transformation,
its energy-momentum tensor changes, as it contains both the metric and its
inverse (through $X$). Transforming them cf. Eq. (\ref{KS}) and $\Tilde{g}%
^{ab}=g^{ab}-\lambda l^{a}l^{b}$ leads to the Kerr--Schild transformed
kinetic term%
\begin{equation}
\Tilde{X}=X+\lambda X^{(1)}\ ,  \label{Xexp}
\end{equation}%
with 
\begin{equation}
X^{(1)}=\dfrac{1}{2}(D\phi )^{2}\ .  \label{X1}
\end{equation}%
The Kerr--Schild transformed energy-momentum tensor is%
\begin{equation}
\tilde{T}_{ab}=F_{\tilde{X}}(\phi ,\tilde{X})\nabla _{a}\phi \nabla _{b}\phi
+\tilde{g}_{ab}F(\phi ,\tilde{X})~.  \label{Tabtilde}
\end{equation}%
When $X^{\left( 1\right) }=0$, the sole change in the energy-momentum tensor
appears through $\tilde{g}_{ab}$. This is possible if the k-essence is
constant along the integral curves of the null congruence. We will discuss
this special case at the end of the paper. In what follows, we concentrate
to the generic case, when $X^{\left( 1\right) }\neq 0$.

\section{K-essence sourcing Kerr--Schild spacetimes}

\subsection{Infinitesimal Kerr--Schild maps}

Until now the parameter $\lambda $ was arbitrary. In this subsection we
assume it is small, hence both functions appearing in the Kerr--Schild
transformed energy-momentum tensor of the k-essence can be expanded in power
series as 
\begin{equation}
F(\phi ,\Tilde{X})=\sum_{j=0}^{\infty }\dfrac{F_{X^{j}}(\phi ,X)}{j!}\left(
\lambda X^{(1)}\right) ^{j}\ ,  \label{F pert}
\end{equation}%
\begin{equation}
F_{\Tilde{X}}(\phi ,\Tilde{X})=\sum_{j=0}^{\infty }\dfrac{F_{X^{j+1}}(\phi
,X)}{j!}\left( \lambda X^{(1)}\right) ^{j}\   \label{F_X pert}
\end{equation}%
(where $F_{X^{j}}$ denotes the $j^{\text{th}}$ derivative of $F$ with
respect to $X$). Hence, for small $\lambda $ and employing Eqs. (\ref{KS}), (%
\ref{Xexp}), (\ref{F pert}), (\ref{F_X pert}) the leading order is given by
the contribution of the seed spacetime: 
\begin{equation}
T_{ab}^{(0)}=T_{ab}\ ,
\end{equation}%
while 
\begin{equation}
T_{ab}^{(k)}=\dfrac{1}{k!}\left( g_{ab}+\nabla _{a}\phi \nabla _{b}\phi 
\dfrac{\partial }{\partial X}\right) F_{X^{k}}(X^{(1)})^{k}+\dfrac{1}{(k-1)!}%
l_{a}l_{b}F_{X^{k-1}}(X^{(1)})^{k-1}\   \label{Tk}
\end{equation}%
holds for any integer $k\geq 1$.

As proven in Ref. \cite{GAL}, the transformed energy-momentum tensor
satisfies the field equations only when all contributions with $k\geq 4$
vanish (this is a generic statement, applying for any $\lambda $, including
small values). Furthermore, if the congruence $l^{a}$ is autoparallel, the $%
k=3$ contribution is also zero. These conditions are expected to seriously
constrain the functional form of the free function $F(\phi ,X)$.

First we prove the following

\emph{Theorem 1.} If $T_{ab}^{(4)}=0$, then for all $k\geq 5$ the
contributions $T_{ab}^{(k)}$ also vanish.

\emph{Proof: }We prove this by induction. Assume that the statement is true
until $k-1\geq 4$, thus 
\begin{gather}
T_{ab}^{(k-1)}=\dfrac{1}{(k-1)!}\left( g_{ab}+\nabla _{a}\phi \nabla
_{b}\phi \dfrac{\partial }{\partial X}\right) F_{X^{k-1}}(X^{(1)})^{k-1} 
\notag \\
+\dfrac{1}{(k-2)!}l_{a}l_{b}F_{X^{k-2}}\left( X^{(1)}\right) ^{k-2}=0~.
\label{Tkm1}
\end{gather}%
Contracting $T_{ab}^{(k-1)}$ with $l^{b}$ and exploring the expression (\ref%
{X1}) for the nonvanishing $X^{\left( 1\right) }$, 
\begin{equation}
l_{a}F_{X^{k-1}}+\sqrt{2X^{(1)}}\nabla _{a}\phi F_{X^{k}}=0\   \label{k-1 l}
\end{equation}%
emerges. Further contracting with $l^{a}$ gives $F_{X^{k}}=0$, which in turn
implies $F_{X^{k-1}}=0$ through Eq. (\ref{k-1 l}). (Hence, the vanishing of $%
T_{ab}^{\left( k-1\right) }$ also implies $F_{X^{k-2}}=0$.) Then%
\begin{equation}
T_{ab}^{(k)}=\dfrac{1}{k!}\nabla _{a}\phi \nabla _{b}\phi
F_{X^{k+1}}(X^{(1)})^{k}~,
\end{equation}%
however $F_{X^{k+1}}$ is also zero, as it is the derivative of a function
vanishing for any $X$. Q.E.D.

Next, let us see, under what conditions would $T_{ab}^{\left( 4\right) }$
vanish. From the proof of Theorem 1, it is immediate to see that it is
equivalent to imposing $F_{X^{3}}=F_{X^{4}}=F_{X^{5}}=0$. This is solved by
quadratic functions of $X$:%
\begin{equation}
F\left( \phi ,X\right) =A\left( \phi \right) X^{2}+B\left( \phi \right)
X-V\left( \phi \right) ~,  \label{Fquad}
\end{equation}%
with \thinspace $A,B,V$ arbitrary functions of $\phi $.

In summary, Kerr--Schild spacetimes with infinitesimal parameter $\lambda $
are solutions of the Einstein equations sourced by k-essence with quadratic
Lagrangian.

\subsection{Finite Kerr--Schild maps}

Let us now ignore that the quadratic form of the k-essence Lagrangian was
derived for infinitesimal Kerr--Schild maps and investigate such maps with
finite parameter for spacetimes sourced by quadratic k-essence. The kinetic
term transforms under such maps with arbitrary $\lambda $ according to Eq. (%
\ref{Xexp}), such that the function $F$ in the Lagrangian becomes 
\begin{equation}
F\left( \phi ,\tilde{X}\right) =A\left( \phi \right) X^{2}+B\left( \phi
\right) X-V\left( \phi \right) +\left[ 2A\left( \phi \right) X+B\left( \phi
\right) \right] \lambda X^{(1)}+A\left( \phi \right) \left( \lambda
X^{(1)}\right) ^{2}~.
\end{equation}%
This agrees with the expression obtained from the expansion (\ref{F pert})
for $F$ quadratic in $X$, confirming its validity for large $\lambda $. The
same conclusion can also be reached by realizing that the convergence radius
of the series expansion is infinite due to the vanishing derivatives.

Therefore, we reached the conclusion that k-essence models with Lagrangian
quadratic in the kinetic term source spacetimes of Kerr--Schild type.

\subsection{Autoparallel null congruence}

The situation is further simplified by requiring the Kerr--Schild null
congruence to be autoparallel, $Dl^{a}\propto l^{a}$. In this case $%
T_{ab}^{(3)}=0$ should also be imposed. We note that the proof of the
Theorem 1 also holds for $k-1=3$, implying $F_{X^{2}}=0$. Therefore, in the
case of autoparallel Kerr--Schild congruences, the k-essence Lagrangian
should be linear in the kinetic term,%
\begin{equation}
F\left( \phi ,X\right) =B\left( \phi \right) X-V\left( \phi \right) ~,
\label{Flin}
\end{equation}%
in order to source Kerr--Schild spacetimes.

\section{Condition for the solution of the linearized system to be exact}

In this section we explore the requirement of the linearized solutions to
also be exact. We start with the autoparallel case, then proceed to the
generic case.

\subsection{Autoparallel Kerr--Schild congruences}

Inserting the linear and quadratic contributions 
\begin{equation}
T_{ab}^{(1)}=l_{a}l_{b}F+F_{X^{2}}X^{(1)}\nabla _{a}\phi \nabla _{b}\phi
+g_{ab}F_{X}X^{(1)}\   \label{T1}
\end{equation}%
and%
\begin{equation}
T_{ab}^{(2)}=\dfrac{1}{2}\left( g_{ab}+\nabla _{a}\phi \nabla _{b}\phi 
\dfrac{\partial }{\partial X}\right)
F_{X^{2}}(X^{(1)})^{2}+l_{a}l_{b}F_{X}(X^{(1)})\   \label{T2}
\end{equation}%
of the energy-momentum tensor into the condition (\ref{cond}), we obtain%
\begin{equation}
\left( g_{ab}+\nabla _{a}\phi \nabla _{b}\phi \dfrac{\partial }{\partial X}%
\right) F_{X^{2}}X^{(1)}=2F_{X^{2}}\left( D\phi \right) l_{(a}\nabla
_{b)}\phi \ .
\end{equation}%
This is automatically solved by the k-essence Lagrangian linear in $X$
(implying $F_{X^{2}}=0$).

\subsection{Unicity}

\emph{Theorem 2.} The solution of the linearized equation becomes exact only
if the Kerr--Schild congruence is autoparallel (hence, the k-essence
Lagrangian is linear in the kinetic term).

\emph{Proof: }Let us assume, that the null congruence $l^{a}$ is generic,
rather than autoparallel. Then, the condition (\ref{cond}) for the linear
solution to become exact is replaced by the statement of Theorem 1 of Ref 
\cite{GAL}, giving%
\begin{eqnarray}
T_{ab}^{\left( 3\right) } &=&-\frac{3}{4}l_{a}l_{b}\left(
Dl^{c}Dl_{c}\right) ~,  \label{T3gen} \\
2T_{ab}^{\left( 2\right) } &=&2l_{(a}T_{b)c}^{\left( 1\right) }l^{c}-\frac{1%
}{2}g_{ab}\left( Dl^{c}Dl_{c}\right) +Dl_{a}Dl_{b}-l_{a}l_{b}\left( \nabla
_{c}Dl^{c}\right)  \notag \\
&&+l_{(a}\left[ Dl_{b)}\left( \nabla _{c}l^{c}\right) +DDl_{b)}+\left(
\nabla _{b)}l_{c}-2\nabla _{\left\vert c\right\vert }l_{b)}\right) Dl^{c}%
\right] ~.  \label{T2gen}
\end{eqnarray}%
The k-essence with quadratic Lagrangian in $X$, Eq. (\ref{Fquad}) has all $%
T_{ab}^{(k\geq 4)}=0$ (therefore able to generate Kerr--Schild type
solutions) and 
\begin{equation}
T_{ab}^{(3)}=\dfrac{1}{4}l_{a}l_{b}A\left( \phi \right) (D\phi )^{4}\ .
\end{equation}%
Comparison with Eq. (\ref{T3gen}) gives the coefficient of the quadratic
contribution: 
\begin{equation}
A\left( \phi \right) =-\frac{3\left( Dl^{c}Dl_{c}\right) }{(D\phi )^{4}}~.
\label{Aphi}
\end{equation}%
The first two expansion coefficients of the energy-momentum tensor, Eqs. (%
\ref{T1}) and (\ref{T2}) are%
\begin{eqnarray}
T_{ab}^{(1)} &=&-\frac{3\left( Dl^{c}Dl_{c}\right) }{(D\phi )^{4}}\left[
l_{a}l_{b}X^{2}+(D\phi )^{2}\left( g_{ab}X+\nabla _{a}\phi \nabla _{b}\phi
\right) \right]  \notag \\
&&+B\left( \phi \right) \left( l_{a}l_{b}X+g_{ab}\dfrac{(D\phi )^{2}}{2}%
\right) -V\left( \phi \right) l_{a}l_{b}
\end{eqnarray}%
and%
\begin{equation}
T_{ab}^{(2)}=-\frac{3\left( Dl^{c}Dl_{c}\right) }{4}\left( g_{ab}+l_{a}l_{b}%
\frac{4X}{(D\phi )^{2}}\right) +B\left( \phi \right) l_{a}l_{b}\frac{(D\phi
)^{2}}{2}\ .
\end{equation}%
Inserting the latter and 
\begin{equation}
2l_{(a}T_{b)c}^{(1)}l^{c}=-\frac{6\left( Dl^{c}Dl_{c}\right) }{(D\phi )^{2}}%
\left( l_{a}l_{b}X+l_{(a}\nabla _{b)}\phi D\phi \right) +B\left( \phi
\right) l_{a}l_{b}(D\phi )^{2}
\end{equation}%
into Eq. (\ref{T2gen}) leads to the condition%
\begin{eqnarray}
6\frac{l_{(a}\nabla _{b)}\phi }{D\phi }\left( Dl^{c}Dl_{c}\right)
&=&Dl_{a}Dl_{b}+g_{ab}\left( Dl^{c}Dl_{c}\right) -l_{a}l_{b}\left( \nabla
_{c}Dl^{c}\right)  \notag \\
&&+l_{(a}\left[ Dl_{b)}\left( \nabla _{c}l^{c}\right) +DDl_{b)}+\left(
\nabla _{b)}l_{c}-2\nabla _{\left\vert c\right\vert }l_{b)}\right) Dl^{c}%
\right] ~.  \label{cond2}
\end{eqnarray}%
It is immediate to check, that in the autoparallel case this becomes an
identity, confirming our previous finding.

For generic null congruences, the trace of Eq. (\ref{cond2}) gives $%
l^{b}DDl_{b}=0$, which can be rewritten as $Dl^{b}Dl_{b}=0$, hence $Dl^{a}$
null. Then, however, through Eq. (\ref{Aphi}) $A=0$. Thus, we have proven,
that in order the linearized Kerr--Schild solution to also be exact, the
k-essence Lagrangian should be linear in the kinetic term.

We complete the proof by exploring the condition of $Dl^{a}$ being null
vector. Beside the autoparallel case $Dl^{a}=\alpha l^{a}$, already
discussed, the other possibility would be $Dl^{a}=\beta k^{a}$, with $k^{a}$
the second null vector of a pseudoorthonormal base (with property $%
k^{a}l_{a}=-1$). In this case, denoting $\delta =k^{c}\nabla _{c}$, Eq. (\ref%
{cond2}) reduces to%
\begin{eqnarray}
0 &=&\beta ^{2}k_{a}k_{b}-l_{a}l_{b}\left( \beta \nabla _{c}k^{c}+\delta
\beta \right)  \notag \\
&&+l_{(a}\left[ k_{b)}\left( \beta \nabla _{c}l^{c}+D\beta \right) +\beta
Dk_{b)}+\beta \left( \nabla _{b)}l_{c}-2\nabla _{\left\vert c\right\vert
}l_{b)}\right) k^{c}\right] ~.  \label{cond3}
\end{eqnarray}%
Its $l^{a}l^{b}$ projection shows $\beta =0$, which renders Eq. (\ref{cond3}%
) into an identity. Therefore, the only surviving possibility is $l^{a}$
being autoparallel. Q.E.D.

\section{Constant k-essence along the integral curves of the Kerr--Schild
null congruence}

For completeness, we also discuss the special case $D\phi =0$, implying $%
X^{\left( 1\right) }=0$. In this case the k-essence is constant along the
integral curves of the Kerr--Schild null congruence and the energy-momentum
tensor changes exclusively due to its dependence on $\tilde{g}_{ab}$:%
\begin{equation}
\tilde{T}_{ab}=T_{ab}+\lambda l_{a}l_{b}F(\phi ,X)~.
\end{equation}%
With only $T_{ab}^{\left( 1\right) }\neq 0$ in the expansion, Eq. (\ref%
{T3gen}) gives $Dl^{c}$ null, while Eq. (\ref{T2gen}) simplifies to 
\begin{eqnarray}
0 &=&Dl_{a}Dl_{b}-l_{a}l_{b}\left( \nabla _{c}Dl^{c}\right)  \notag \\
&&+l_{(a}\left[ Dl_{b)}\left( \nabla _{c}l^{c}\right) +DDl_{b)}+\left(
\nabla _{b)}l_{c}-2\nabla _{\left\vert c\right\vert }l_{b)}\right) Dl^{c}%
\right] ~,  \label{cond4}
\end{eqnarray}%
a condition purely on the null congruence ($F$ dropped out).

For autoparallel congruences $Dl^{a}=\alpha l^{a}$ the condition (\ref{cond4}%
) becomes an identity. For the alternative case $Dl^{a}=\beta k^{a}$ it gives%
\begin{eqnarray}
0 &=&\beta ^{2}k_{a}k_{b}-l_{a}l_{b}\left( \beta \nabla _{c}k^{c}+\delta
\beta \right)  \notag \\
&&+l_{(a}\left[ \beta k_{b)}\left( \nabla _{c}l^{c}\right) +\beta
Dk_{b)}+k_{b)}D\beta +\beta \left( k^{c}\nabla _{b)}l_{c}-2\delta
l_{b)}\right) \right] ~,
\end{eqnarray}%
its $l^{a}l^{b}$ projection implying $\beta =0$ leads to another identity.

With this, we have proven:

\emph{Theorem 3. }For k-essence fields constant along the integral curves of
the autoparallel null Kerr--Schild congruence the solutions of the
linearized Einstein equations also solve the exact equations, with no
further restriction on the functional form of the k-essence Lagrangian.

\section{On the physical interpretation of Kerr--Schild maps}

\subsection{How are black hole properties affected by Kerr--Schild maps}

It is interesting to consider, how the Kerr--Schild map transforms the
characteristics of spacetime, as it affects null geodesics, hence causality,
horizon location, light deflection, gravitational lensing, and black hole
shadows. The latter gained particular interest in light of the Event Horizon
Telescope observations of the M87* and Sagittarius A* supermassive black
holes \cite{EH1,EH2}.

The Kerr--Schild map changes the light cone in each point, by modifying it
everywhere, except along one conserved direction generated by the
Kerr--Schild congruence. For Kerr black holes with mass $M$ and rotation
parameter $a$ the deformation caused by a Kerr--Schild map can be easily
visualized. The Kerr metric in Kerr--Schild coordinates ($t^{\prime },x,y,z$%
) is 
\begin{equation}
\tilde{g}_{ab}=\eta _{ab}+Hl_{a}^{\prime }l_{b}^{\prime }~,
\end{equation}%
with 
\begin{equation}
H=\frac{2Mr^{3}}{r^{4}+a^{2}z^{2}}~,\quad l_{a}^{\prime }=\left( 1,\frac{%
rx+ay}{r^{2}+a^{2}},\frac{ry-ax}{r^{2}+a^{2}},\frac{z}{r}\right) ~,
\label{Hl}
\end{equation}%
where the constant $r$ surfaces are ellipsoidal, emerging from the null
condition $l^{\prime a}l_{a}^{\prime }=0$ as 
\begin{equation}
\frac{x^{2}}{r^{2}+a^{2}}+\frac{y^{2}}{r^{2}+a^{2}}+\frac{z^{2}}{r^{2}}=1~.
\end{equation}%
Note that the form $Hl_{a}^{\prime }l_{b}^{\prime }$ can be obtained from $%
\lambda l_{a}l_{b}$ by reparametrizing the null congruence. We also remark
that for any decomposition $M=M_{1}+M_{2}$ (and denoting $%
H_{i}=2M_{i}r^{3}/\left( r^{4}+a^{2}z^{2}\right) $, with $i=1,2$), the Kerr
metric in the Kerr--Schild form can be decomposed in two equivalent ways: 
\begin{equation}
\tilde{g}_{ab}=\eta _{ab}+\left( H_{1}l_{a}^{\prime }l_{b}^{\prime
}+H_{2}l_{a}^{\prime }l_{b}^{\prime }\right) =\left( \eta
_{ab}+H_{1}l_{a}^{\prime }l_{b}^{\prime }\right) +H_{2}l_{a}^{\prime
}l_{b}^{\prime }~.
\end{equation}%
The first decomposition is the initial interpretation of a Kerr--Schild map
acting on the flat seed spacetime, transforming it into a Kerr spacetime
with mass $M=M_{1}+M_{2}$. The second decomposition represents a
Kerr--Schild map from a Kerr seed spacetime with mass $M_{1}$ to another
Kerr spacetime with mass $M$. Hence, one can interpret the Kerr--Schild map
acting on a Kerr black hole as a simple increase of the mass. This means,
that deflection of light will increase, lensing will be amplified, and the
radius of the black hole shadow will increase. A further remark concerns the
null direction unaffected by the Kerr--Schild map. In the nonrotating case $%
a=0$ the Kerr--Schild congruence from Eq. (\ref{Hl}) has purely radial
spatial projection ($x/r,y/r,z/r$). In the rotating case $a\neq 0$ these
projections will point perpendicularly to the ellipsoidal surface ($%
l^{\prime a}dl_{a}^{\prime }=0$ holds). Therefore the Kerr--Schild map is
conserving the symmetries and is not expected to change the shape of the
black hole shadow.

\subsection{The scalar field inside and outside a black\ hole}

In this subsection we qualitatively discuss the modifications induced by a
k-essence scalar field of the type allowed by our Theorems on black holes.
It is well known, that a scalar with a timelike gradient is equivalent to a
perfect fluid \cite{Faraoni}, at least at the nonperturbative level. By
contrast, when the spacetime is static and spherically symmetric, the scalar
field depends only on the radial coordinate, hence its gradient is
spacelike. In this case the scalar energy-momentum tensor (\ref{Tab}) is
equivalent to an imperfect fluid with its tangential pressure equal to the
negative of its energy density \cite{ScalarFluidEquiv}.

These properties can easily be seen by inserting the metric decomposition 
\begin{equation}
g_{ab}=-n_{a}n_{b}+m_{a}m_{b}+h_{ab}~,  \label{decomp1}
\end{equation}%
(where $n^{a}$ and $m^{a}$ are an orthornormal pair, also normal to $h_{ab}$%
) into the energy-momentum tensor (\ref{Tab}), yielding%
\begin{equation}
T_{ab}=\left( 2XF_{X}-F\right) n_{a}n_{b}+F\left( m_{a}m_{b}+h_{ab}\right) ~,
\label{TabT}
\end{equation}%
when the scalar gradient is timelike, $\nabla _{a}\phi \propto X^{1/2}n_{a}$
or%
\begin{equation}
T_{ab}=\left( F-2XF_{X}\right) m_{a}m_{b}+F\left( -n_{a}n_{b}+h_{ab}\right)
~,  \label{TabS}
\end{equation}%
when the scalar gradient is spacelike $\nabla _{a}\phi \propto \left(
-X\right) ^{1/2}m_{a}$.

In the spherically symmetric case the scalar gradient points along $\partial
/\partial r$. However this direction transitions from spacelike outside the
horizon to timelike inside the black hole.

\subsubsection{Inside the horizon}

The scalar field trapped inside the horizon mimics a perfect fluid with
energy density $\rho _{\mathrm{in}}=2XF_{X}-F$ and isotropic pressure $p_{%
\mathrm{in}}=F$. For the quadratic case (\ref{Fquad}) the energy density and
isotropic pressure become%
\begin{eqnarray}
\rho _{\mathrm{in}} &=&3A\left( \phi \right) X^{2}+B\left( \phi \right)
X+V\left( \phi \right) ~,  \notag \\
p_{\mathrm{in}} &=&A\left( \phi \right) X^{2}+B\left( \phi \right) X-V\left(
\phi \right) \ .  \label{pinrhoin}
\end{eqnarray}%
These contribute to the gravitational attraction of the black hole in the
same manner like stellar matter (through the Tolman--Oppenheimer--Volkoff
equation, which however, in this case appears as an integro-differential
equation, with the mass function defined as an integral in terms of the
functions \thinspace $A,B,V$). The equation of state for the scalar results
in%
\begin{equation}
w_{\mathrm{in}}=\frac{p_{\mathrm{in}}}{\rho _{\mathrm{in}}}=1-\frac{2\left(
AX^{2}+V\right) }{3AX^{2}+BX+V}~.  \label{win}
\end{equation}%
When the potential dominates, $w_{\mathrm{in}}\approx -1$, thus the scalar
mimics dark energy.

The adiabatic sound speed (the propagation velocity of the scalar field
perturbations) squared is%
\begin{equation}
c_{s,\mathrm{in}}^{2}=\left. \frac{dp_{\mathrm{in}}}{d\rho _{\mathrm{in}}}%
\right\vert _{s/n}=\left. \frac{p_{\mathrm{in},X}}{\rho _{\mathrm{in},X}}%
\right\vert _{s/n}=\frac{2AX+B}{6AX+B}=1-\frac{4AX}{6AX+B}~,
\end{equation}%
where $s=S/V$ is the entropy density and $n=N/V$ the particle number
density. We have explored that $d(s/n)=0$ yields $d\phi =0$ (a simple
realization of this being $\phi =s/n$). The condition $d\phi =0$ signifies
that while the variations $dp$ and $d\rho $ allow for arbitrary variations
in $X$, they are such that $\phi $ should stay constant \cite{sound}.

The pairs of functions $A,B$ yielding Laplacian instability regimes with $%
c_{s,\mathrm{in}}^{2}<0$ should be excluded.

\subsubsection{ Outside the horizon}

In this case the scalar is equivalent to an anisotropic fluid with energy
density equal to the tangential pressures $\rho _{\mathrm{out}}=p_{\mathrm{%
out}}^{t}=F$ and radial pressure $p_{\mathrm{out}}^{r}=F-2XF_{X}$, which in
the quadratic case (\ref{Fquad}) become%
\begin{eqnarray}
\rho _{\mathrm{out}} &=&p_{\mathrm{out}}^{t}=A\left( \phi \right)
X^{2}+B\left( \phi \right) X-V\left( \phi \right) ~,  \notag \\
p_{\mathrm{out}}^{r} &=&-3A\left( \phi \right) X^{2}-B\left( \phi \right)
X-V\left( \phi \right) ~,
\end{eqnarray}%
resulting in 
\begin{equation}
w_{\mathrm{out}}^{r}=\frac{p_{\mathrm{out}}^{r}}{\rho _{\mathrm{out}}}=-1-%
\frac{2\left( AX^{2}+V\right) }{AX^{2}+BX-V}~,\quad w_{\mathrm{out}}^{t}=%
\frac{p_{\mathrm{out}}^{t}}{\rho _{\mathrm{out}}}=1~.
\end{equation}%
When the potential dominates, $w_{\mathrm{out}}^{r}\approx 1$, thus the
fluid approaches isotropy

\begin{eqnarray}
\left( c_{s,\mathrm{out}}^{r}\right) ^{2} &=&\left. \frac{dp_{\mathrm{out}%
}^{r}}{d\rho _{\mathrm{out}}}\right\vert _{s/n}=\left. \frac{p_{\mathrm{out}%
,X}^{r}}{\rho _{\mathrm{out},X}}\right\vert _{s/n}=~-1-\frac{4AX}{2AX+B} \\
\left( c_{s,\mathrm{out}}^{t}\right) ^{2} &=&\frac{dp_{\mathrm{out}}^{t}}{%
d\rho _{\mathrm{out}}}=1~.
\end{eqnarray}%
Again, any pair of functions $A,B$ yielding radial Laplacian instability
regimes with $\left( c_{s,\mathrm{out}}^{r}\right) ^{2}<0$ should be
excluded. Tangential Laplacian instability regimes do not arise.

\subsection{Cosmological scalar field}

In a\ cosmological setup the scalar field is equivalent to a perfect fluid (%
\ref{TabT}), with energy density and pressure already calculated as (\ref%
{pinrhoin}) and equation of state as (\ref{win}). The scalar field mimics
dark energy (cosmological constant), when the potential dominates. The sole
difference compared to the discussion on the black hole interior arises from
the fact that in this case the scalar depends on time (instead of the radial
coordinate). Hence, the adiabatic speed of sound squared reads 
\begin{equation}
c_{s}^{2}=\left. \frac{dp}{d\rho }\right\vert _{s/n}=\left. \frac{p_{X}}{%
\rho _{X}}\right\vert _{s/n}=1-\frac{4AX}{6AX+B}~.
\end{equation}%
A physical requirement to impose on the set of functions $A,B$ is to avoid
the instability regime $c_{s}^{2}<0$.

\subsection{On k-essence Kerr--Schild seed spacetimes}

In order to apply our Theorems, a seed spacetime generated by a nontrivial
scalar field is needed. The simplest such solutions are expected to arise in
highly symmetric situations. However various unicity theorems forbid the
scalar hair for k-essence black holes. Bekenstein ruled out the existence of
stationary, asymptotically flat black holes with scalar hair for canonical
scalar fields (quintessence) \cite{Bek1,Bek2,Bek3}. The generalized
Brans--Dicke theories in the Einstein frame are also contained in this
class, hence its stationary and asymptotically flat black holes have no
scalar hair either \cite{Sotiriou}. Another no-hair theorem for stationary,
asymptotically flat black holes in a more generic class of k-essence models
was provided by Graham and Jha \cite{GJ}, holding when $F_{X}$ and $\phi
F_{\phi }$ are of opposite and definite signs. No-hair theorems were also
shown to hold for static, asymptotically flat black holes in Horndeski,
Beyond Horndeski, Einstein--scalar--Gauss--Bonnet, and Chern--Simons
theories \cite{Ferreira}.

Due to the host of unicity theorems assuming stationarity and asymptotic
flatness, we expect that our result would be useful in dynamical situations.
Such a scenario could be a black hole with slowly growing scalar hair,
arising either from cosmological evolution, or due to the slow motion of the
black hole within an asymptotic spatial gradient in the scalar field \cite%
{Jacobson}. Pairs of inspiralling black holes of this kind emit dipole
radiation, first constrained by observations on the quasar OJ287 in Ref. 
\cite{Burgess}. Another model could be a Vaidya-type radiating solution with
dynamical horizon, as discussed for a Dirac--Born--Infeld model in Ref. \cite%
{Manna3}. Perhaps the most interesting would be considering wavelike
behaviours. For example, Einstein--Rosen cylindrical waves were derived for
Brans--Dicke theories (which in Einstein frame fit into our framework) in
Ref. \cite{AkyarDelice}, including standing wave, solitonic wave, and pulse
wave type particular solutions. For any such time evolving solution obtained
from the k-essence Lagrangian linear in $X$, a Kerr--Schild spacetime would
emerge as the solution of the Einstein equation linearized in $\lambda $
instead of dealing with the full set of Einstein equations. Exploring such
possibilities require further investigations beyond the scope of this paper.

\section{Concluding Remarks}

Most of the physically interesting solutions of the Einstein equations,
including black holes and radiation fields are of Kerr--Schild type. The
generating technique of such solutions is quite elegant geometrically: the
Kerr--Schild map modifies the metric through the addition of a diad of null
vectors, therefore changing the light-cones in such a way, that in each
point a single null generator of the cone is left unmodified. One of the
most fascinating properties of such Kerr--Schild spacetimes in vacuum is
that any solution of the linearized Einstein equation can be propelled into
an exact solution of the full nonlinear Einstein equations by simply
increasing the expansion parameter to finite values \cite{Xanthopoulos}. The
conditions for generalising this property to the case, when matter sources
are present are also known \cite{GAL}.

In the present paper we have investigated whether this property of
Kerr--Schild maps holds for minimally coupled k-essence scalar-tensor
theories. First, we proved that the Lagrangian of the k-essence should be at
most quadratic in the kinetic term in order to source a Kerr--Schild type
spacetime. In a cosmological context, such models include dilatonic ghost
condensate \cite{dilatonicghost} and unified models of dark energy and dark
matter \cite{Scherrer}.

This is reduced to a linear dependence $\mathfrak{L}_{\phi }=\sqrt{-%
\mathfrak{g}}\left[ B\left( \phi \right) X-V\left( \phi \right) \right] $,
when the Kerr--Schild null congruence is autoparallel. Generic junction
conditions and the generalization of Lanczos equation were derived \cite{RG}
for such k-essence fields, which also include the quintessence models in the
particular case $B=1$.

Further, we proved a unicity theorem for k-essence. The solutions of the
Einstein equations linearized in Kerr--Schild type perturbations also solve
the full nonlinear system of Einstein equations, when the k-essence is given
by a linear Lagrangian in the kinetic term. The proof of the theorem omitted
one special case, of a k-essence everywhere constant along the integral
curves of the Kerr--Schild congruence. We proved that such a k-essence field
also shares the property of sourcing perturbative Kerr--Schild spacetimes
which are exact, without the need to restrict in any way the functional form
of its Lagrangian.

\begin{acknowledgments}
B.J. is grateful for the support received during his internship at HUN-REN
Wigner RCP.
\end{acknowledgments}


\begin{thebibliography}{99}
\bibitem{ExactSol} H. Stephani, D. Kramer, M. Maccallum, C. Hoenselaers, and
E. Herlt, Exact Solutions of Einstein's Field Equations, Cambridge
University Press, 485 (2023)

\bibitem{KS1} L. \'{A}. Gergely and Z. Perj\'{e}s, Kerr-Schild metrics
revisited I. The ground state, J. Math. Phys. \textbf{35}, 2438 (1994)
[arXiv:gr-qc/0203088].

\bibitem{KS2} L. \'{A}. Gergely and Z. Perj\'{e}s, Kerr-Schild metrics
revisited II. The complete vacuum solution, J. Math. Phys. \textbf{35}, 2448
(1994) [arXiv:gr-qc/0203090].

\bibitem{KSB0} L. \'{A}. Gergely and Z. Perj\'{e}s, Vacuum Kerr-Schild
metrics generated by nontwisting congruences, Ann. Physik \textbf{3}, 609
(1994) [arXiv:gr-qc/0203091].

\bibitem{Xanthopoulos} B. C. Xanthopoulos, Exact vacuum solutions of
Einstein's equation from linearized solutions, J. Math. Phys \textbf{19},
1607 (1978).

\bibitem{GAL} L. \'{A}. Gergely, Linear Einstein equations and Kerr-Schild
maps, Class. Quantum Grav. \textbf{19,} 2515 (2002) [arXiv:gr-qc/0203101].

\bibitem{TestingGRtopical} E. Berti et al., Testing general relativity with
present and future astrophysical observations, Class. Quantum Grav. \textbf{%
32}, 243001 (2015) [arXiv:1501.07274 [gr-qc]].

\bibitem{GursesVT} M. G\"{u}rses and \c{C}. \c{S}ent\"{u}rk, A Modified
Gravity Theory: Null Aether, Commun. Theor. Phys. \textbf{71}, 312 (2019)
[arXiv:1604.02266 [gr-qc]].

\bibitem{GursesH} M. G\"{u}rses, Y. Heydarzade, and \c{C}. \c{S}ent\"{u}rk,
Kerr--Schild--Kundt metrics in generic gravity theories with modified
Horndeski couplings, Eur. Phys. J. C \textbf{81}, 1147 (2021)
[arXiv:2109.01244 [gr-qc]].

\bibitem{Horndeski} G. W. Horndeski, Second-order scalar-tensor field
equations in a four-dimensional space, International Journal of Theoretical
Physics. \textbf{10(6)}, 363-384 (1974).

\bibitem{Deffayet} C. Deffayet, X. Gao, D. A. Steer, and G. Zahariade, From
k-essence to generalized Galileons, Phys. Rev. D. \textbf{84}, 064039 (2011)
[arXiv:1103.3260 [hep-th]].

\bibitem{Multimessenger} LIGO Scientific and Virgo Collaborations, Fermi
Gamma-ray burst monitor, and INTEGRAL, Gravitational Waves and Gamma-Rays
from a Binary Neutron Star Merger: GW170817 and GRB170817A, Astrophys. J.
Lett. \textbf{848}, L13 (2017) [arXiv:1710.05834 [astro-ph.HE]].

\bibitem{GWc1} T. Baker, E. Bellini, P. G. Ferreira, M. Lagos, J. Noller,
and I. Sawicki, Strong constraints on cosmological gravity from GW170817 and
GRB 170817A, Phys. Rev. Lett. \textbf{119}, 251301 (2017) [arXiv:1710.06394
[astro-ph.CO]].

\bibitem{GWc2} J. M. Ezquiaga and M. Zumalac\'{a}rregui, Dark Energy after
GW170817: Dead ends and the road ahead, Phys. Rev. Lett. \textbf{119},
251304 (2017) [arXiv:1710.05901 [astro-ph.CO]].

\bibitem{GWc3} P. Creminelli and F. Vernizzi, Dark Energy after GW170817 and
GRB170817A, Phys. Rev. Lett. \textbf{119}, 251302 (2017) [arXiv:1710.05877
[astro-ph.CO]].

\bibitem{GWc4} J. Sakstein and B. Jain, Implications of the Neutron Star
Merger GW170817 for Cosmological Scalar-Tensor Theories, Phys. Rev. Lett. 
\textbf{119}, 251303 (2017) [arXiv:1710.05893 [astro-ph.CO]].

\bibitem{KGB} C. Deffayet, O. Pujolas, I. Sawicki, and A. Vikman, Imperfect
Dark Energy from Kinetic Gravity Braiding, JCAP \textbf{1010}, 026 (2010)
[arXiv:1008.0048 [hep-th]].

\bibitem{kinfl} C. Armendariz-Picon, T. Damour, and V. Mukhanov,
k-Inflation, Phys. Lett. B458, 209 (1999) [arXiv:hep-th/9904075].

\bibitem{lateDE} C. Armendariz-Picon, V. Mukhanov, and P. J. Steinhardt, A
Dynamical Solution to the Problem of a Small Cosmological Constant and
Late-Time Cosmic Acceleration, Phys. Rev. Lett. \textbf{85}, 4438 (2000)
[arXiv:astro-ph/0004134].

\bibitem{kess} C. Armendariz-Picon, V. F. Mukhanov, and P. J. Steinhardt,
Essentials of k-essence Phys. Rev. D \textbf{63}, 103510 (2001)
[arXiv:astro-ph/0006373].

\bibitem{DDE} E. A. Lim, I. Sawicki, and A. Vikman, Dust of Dark Energy,
JCAP \textbf{2010}, 05:012, (2010) [arXiv:1003.5751 [astro-ph.CO]].

\bibitem{Speeding} O. Luongo and M. Muccino, Speeding up the universe using
dust with pressure, Phys. Rev. D \textbf{98}, 103520 (2018)
[arXiv:1807.00180 [gr-qc]].

\bibitem{Healing} R. D'Agostino, O. Luongo, and M. Muccino, Healing the
cosmological constant problem during inflation through a unified
quasi-quintessence matter field, Class. Quantum Grav. \textbf{39}, 195014
(2022) [arXiv:2204.02190 [gr-qc]].

\bibitem{Luongo} O. Luongo and T. Mengoni, Generalized K-essence inflation
in Jordan and Einstein frames, Class. Quantum Grav. \textbf{41}, 105006
(2024).

\bibitem{KEBH} R. Akhoury, D. Garfinkle, and R. Saotome, Gravitational
collapse of k-essence, J. High Energy Phys. \textbf{2011, }096, (2011)
[arXiv:1103.0290 [gr-qc]].

\bibitem{Manna1} G. Manna, P. Majumdar, and B. Majumder, k-essence emergent
spacetime as a generalized Vaidya geometry, Phys. Rev. D \textbf{101},
124034 (2020) [arXiv:1909.07224 [gr-qc]].

\bibitem{Manna2} G. Manna, Gravitational collapse for the K-essence emergent
Vaidya spacetime, Eur. Phys. J. C \textbf{80}, 813, (2020) [arXiv:1911.11753
[gr-qc]].

\bibitem{Manna3} B. Majumder, S. Ray, G. Manna, Evaporation of Dynamical
Horizon with the Hawking Temperature in the K-essence Emergent Vaidya
Spacetime, Fortschritte der Physik \textbf{71}, 2300133 (2023)
[arXiv:2007.16053 [gr-qc]].

\bibitem{classific} R. Kase and S. Tsujikawa, Dark energy in Horndeski
theories after GW170817: A review, Int. J. Mod. Phys. D \textbf{28}, 1942005
(2019) [arXiv:1809.08735 [gr-qc]].

\bibitem{EH1} Event Horizon Telescope Collaboration, First M87 Event Horizon
Telescope Results. VI. The Shadow and Mass of the Central Black Hole,
Astrophys. J. Lett. \textbf{875}, L6 (2019) [arXiv:1906.11243].

\bibitem{EH2} Event Horizon Telescope Collaboration, First Sagittarius A*
Event Horizon Telescope Results. I. The Shadow of the Supermassive Black
Hole in the Center of the Milky Way, Astrophys. J. Lett. \textbf{930}, L12
(2022) [arXiv:2311.08680].

\bibitem{Faraoni} V. Faraoni, Correspondence between a scalar field and an
effective perfect fluid, Phys. Rev. D \textbf{85}, 024040 (2012)
[arXiv:1201.1448 [gr-qc]].

\bibitem{ScalarFluidEquiv} C. Gergely, Z. Keresztes, and L. \'{A}. Gergely,
Minimally coupled scalar fields as imperfect fluids, Phys. Rev. D \textbf{102%
}, 024044 (2020) [arXiv:2007.01326 [gr-qc]].

\bibitem{sound} O. F. Piattella, J. C. Fabris, and N. Bili\'{c}, Note on the
thermodynamics and the speed of sound of a scalar field, Class. Quantum
Grav. \textbf{31}, 055006 (2014) [arXiv:1309.4282 [gr-qc]].

\bibitem{Bek1} J. D. Bekenstein, Transcendence of the Law of Baryon-Number
Conservation in Black-Hole Physics, Phys. Rev. Lett. \textbf{28}, 452 (1972).

\bibitem{Bek2} J. D. Bekenstein, Nonexistence of Baryon Number for Static
Black Holes, Phys. Rev. D \textbf{5}, 1239 (1972).

\bibitem{Bek3} J. D. Bekenstein, Nonexistence of Baryon Number for Black
Holes. II, Phys. Rev. D \textbf{5}, 2403 (1972).

\bibitem{Sotiriou} T.P. Sotiriou, Black holes and scalar fields, Class.
Quantum Grav. \textbf{32}, 214002 (2015) [arXiv:1505.00248 [gr-qc].

\bibitem{GJ} A. A. H. Graham and R. Jha, Nonexistence of black holes with
noncanonical scalar fields, Phys. Rev. D \textbf{89}, 084056 (2014)
[arXiv:1401.8203 [gr-qc]].

\bibitem{Ferreira} O. J. Tattersall, P. G. Ferreira, and M. Lagos, Speed of
gravitational waves and black hole hair, Phys. Rev. D \textbf{97}, 084005
(2018) [arXiv:1802.08606 [gr-qc]].

\bibitem{Jacobson} T. Jacobson, Primordial black hole evolution in
tensor-scalar cosmology, Phys.Rev.Lett. \textbf{83}, 2699, (1999) [arXiv:
astro-ph/9905303].

\bibitem{Burgess} M.W. Horbatsch and C.P. Burgess, Cosmic black-hole hair
growth and quasar OJ287, J. Cosmol. Astropart. Phys. \textbf{1205,} 010
(2012) [arXiv:1111.4009 [gr-qc]].

\bibitem{AkyarDelice} L. Akyar and \"{O}. Delice, On generalized
Einstein-Rosen waves in Brans-Dicke theory, Eur. Phys. J. Plus \textbf{129},
226 (2014).

\bibitem{dilatonicghost} F. Piazza and S. Tsujikawa, Dilatonic ghost
condensate as dark energy, J. Cosmol. Astropart. Phys. \textbf{0407}, 004
(2004) [hep-th/0405054].

\bibitem{Scherrer} R. J. Scherrer, Purely kinetic k-essence as unified dark
matter, Phys. Rev. Lett. \textbf{93}, 011301 (2004) [astro-ph/0402316].

\bibitem{RG} B. Racsk\'{o} and L. \'{A}. Gergely, The Lanczos Equation on
Light-Like Hypersurfaces in a Cosmologically Viable Class of Kinetic Gravity
Braiding Theories, Symmetry \textbf{11}, 616 (2019) [arXiv:2006.13247
[gr-qc]].
\end{thebibliography}
\end{document}